\documentclass[review]{elsarticle}

\journal{Nucl. Instrum. Methods Phys. Res.}
\biboptions{sort&compress}

\usepackage[separate-uncertainty=true]{siunitx}
\usepackage[T1]{fontenc}
\usepackage[utf8]{inputenc}
\usepackage[english]{babel}
\usepackage{xspace}
\usepackage{multirow}
\usepackage[colorlinks=true]{hyperref}
\usepackage{lineno}

\newcommand\panda{$\overline{\text{P}}$ANDA\xspace}

\begin{document}

\begin{frontmatter}

\title{An infrared light-guide based target positioning system for operation in a harsh environment}
\author[1]{Falk Schupp\texorpdfstring{\corref{cor1}}{}}
\ead{schupp@uni-mainz.de}
\author[1]{Michael Bölting}
\author[1,2]{Patrick Achenbach\fnref{jlab}}
\author[1]{Sebastian Bleser}
\author[1,2]{Josef Pochodzalla}
\author[1]{Marcell Steinen}
\author{\\~\\ on behalf of the \panda collaboration}
\address[1]{Helmholtz Institute Mainz, GSI Helmholtz Centre for Heavy Ion Research, Darmstadt, Johannes Gutenberg University, 55099 Mainz, Germany}
\address[2]{Institute for Nuclear Physics, Johannes Gutenberg University, 55099 Mainz, Germany}

\fntext[jlab]{Now at Thomas Jefferson National Accelerator Facility,
12000 Jefferson Avenue, Newport News, VA, USA.}

\cortext[cor1]{Corresponding author}

\begin{abstract}
In the \panda experiment's hypernuclear and hyperatom setup, a positioning system for the primary production target is required, which will be located in the center of the solenoid magnet, in ultra-high vacuum, and exposed to high radiation levels. In this work, a prototype for a positioning sensor was built using a bisected light guide for infrared light and a low-priced readout system based on microcontrollers. In contrast to many modern positioning systems that require electronics in direct proximity, this setup has no active electronic components close to the moving parts.

The prototype system was operated with a resolution of better than \SI{5}{\micro\m}, and with a repeatability of better than $\pm$ \SI{18}{\micro\m} in a total of 14\,000 measurements. The demonstrated performance is by far satisfying the positioning requirement of $\pm$ \SI{300}{\micro\m} in the hypernuclear and hyperatom setup at \panda.

\end{abstract}

\end{frontmatter}


\section{Introduction}
%
The large production cross section for hyperons and antihyperons in antiproton induced interactions
provides an unprecedented tool to study strange baryons and antibaryons in nuclear system. The antiProton ANnihilations at DArmstadt ({\panda}) experiment, which is a planned fixed-target experiment at the international Facility for Antiproton and Ion Research (FAIR)~\cite{Barucca_2021}, is the ideal place for such studies. {\panda} is intended to operate at the antiproton storage ring HESR and it will address various aspects of the strong interaction, with strangeness nuclear physics being one of its research pillars~\cite{POCHODZALLA2005430,Lorente2011222,Pochodzalla2008306,SanchezLorente2015421,Singh2016323,Steinen:20225N,HYP2022_refId23}. These studies are made possible by the modular structure of the \panda detector which allows to integrate a dedicated target station, that combines a primary production target with a secondary target to stop the produced hyperons and antihyperons. This paper focuses on the positioning system for the primary targets in the center of the target station.

\section{The target system of the hypernuclear and hyperatom setup}
%
\begin{figure}[htbp]
    \centering
    \includegraphics[width=\columnwidth]{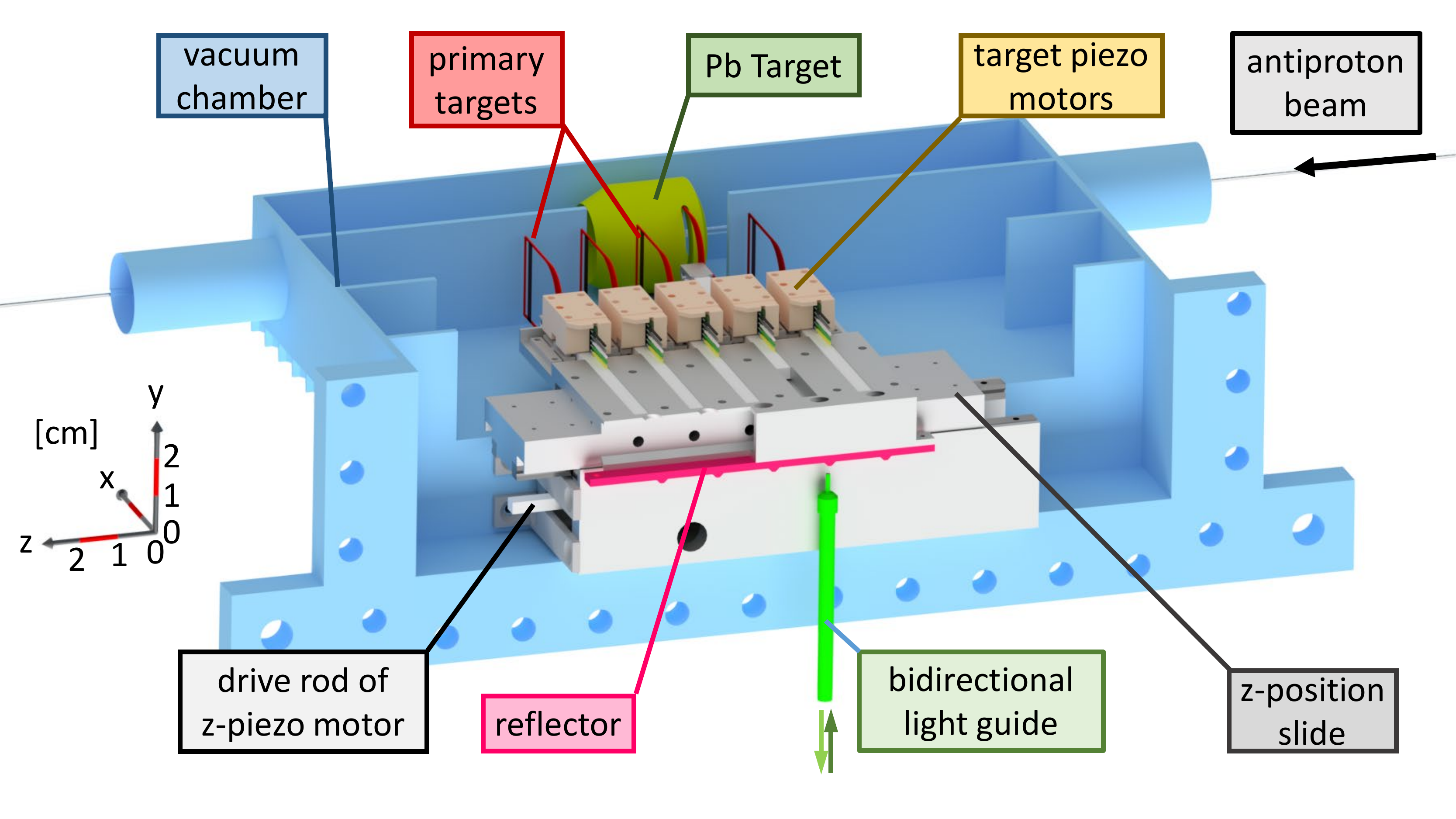}
    \caption{CAD drawing of the target setup indented to be used at \panda. The beam enters from the right and passes through a hollow Pb barrel (green) along its axis. Up to five diamond filament targets are mounted on thin frames (red). A frame can be moved into the barrel through a thin slot by a piezo motor. In case a filament needs to be replaced, its frame will be moved out of the barrel and a slide (gray) will bring another frame into position. The piezo motor for the movement of the slide in $z$-direction is mounted beneath.}
    \label{fig:setup}
\end{figure}
The primary target will be made from a thin carbon filament which will be moved into the halo of the antiproton beam. It will be located inside the \panda solenoid approximately \SI{55}{cm} upstream of the hydrogen target position~\cite{Singh2016323}. For the strangeness nuclear physics measurements, $\Xi^-$ hyperons will be produced in $\overline{p}\,+\,^{12}$C interactions
and subsequently stopped inside the secondary target. In case of the planned $\Xi^-$ hyperatom studies, this secondary target will be made of a Pb barrel surrounding the beam and the primary target.
The primary target system has to meet several requirements:
\begin{itemize}
\item
During the filling of the storage ring with antiprotons, the filament needs to be retracted from the circulating beam. This will happen typically once per hour.
\item
During data taking, the number of antiprotons will decrease monotonously. In order to keep the luminosity constant, the filament has to be moved continuously closer to the center of the beam by up to one millimeter.
\item
Carbon filaments are very filigree and radiation damage during operation could require a replacement.
Therefore, the system must be able to remotely extract a damaged filament, move a replacement into position, and insert it through a thin slit into the beam pipe.
\item
The system must operate in ultra-high vacuum, inside a \SI{2}{T} solenoid magnetic field, and less than \SI{10}{cm} away from the antiproton interaction point.
\item
To optimize the capture probability of the short-lived $\Xi^-$ and to minimize the absorption of gamma rays emitted from hyperatoms or hypernuclei, the system must be as compact as possible.
\end{itemize}
The environmental conditions pose severe restriction on the use of commercially available active electronics for monitoring the movements. Therefore, we developed a system based on piezo motors for positioning and an infrared sensor with light guides for monitoring.
In this system there is no need for active electronics in the proximity of the antiproton interaction point, and it is capable of operating under a wide range of environmental conditions to satisfy the above requirements.
Although we refer in this paper to the target systems of the planned hyperatom and hypernucei measurements at \panda, a similar setup based on piezo motors as driving systems may be used in other experiments where similar conditions apply.

\autoref{fig:setup} shows a CAD drawing of the target setup indented to be used at {\panda} for the study of $\Xi^-$-Pb hyperatoms. The size of the setup is indicated by the coordinate system on the left. The beam enters from the right and passes through a hollow Pb barrel along its axis. The diamond filaments are mounted on thin frames. A frame can be moved into the barrel through a \SI{2}{\mm} wide slot by a piezo motor. In case a filament needs to be replaced, its frame will be moved out of the barrel and a slide can be moved to an alternative position to move in another frame. The slide movement in $z$-position is realized by a piezo motor mounted beneath.

The overall thickness of the frame is about \SI{500}{\micro\m}. Considering the possible side clearance of the moving rod and mechanical uncertainties, the system is required to reliably position a frame within bounds of about \SI{\pm300}{\micro\m}.

\begin{figure}[htbp]
    \centering
    \includegraphics[width=\columnwidth]{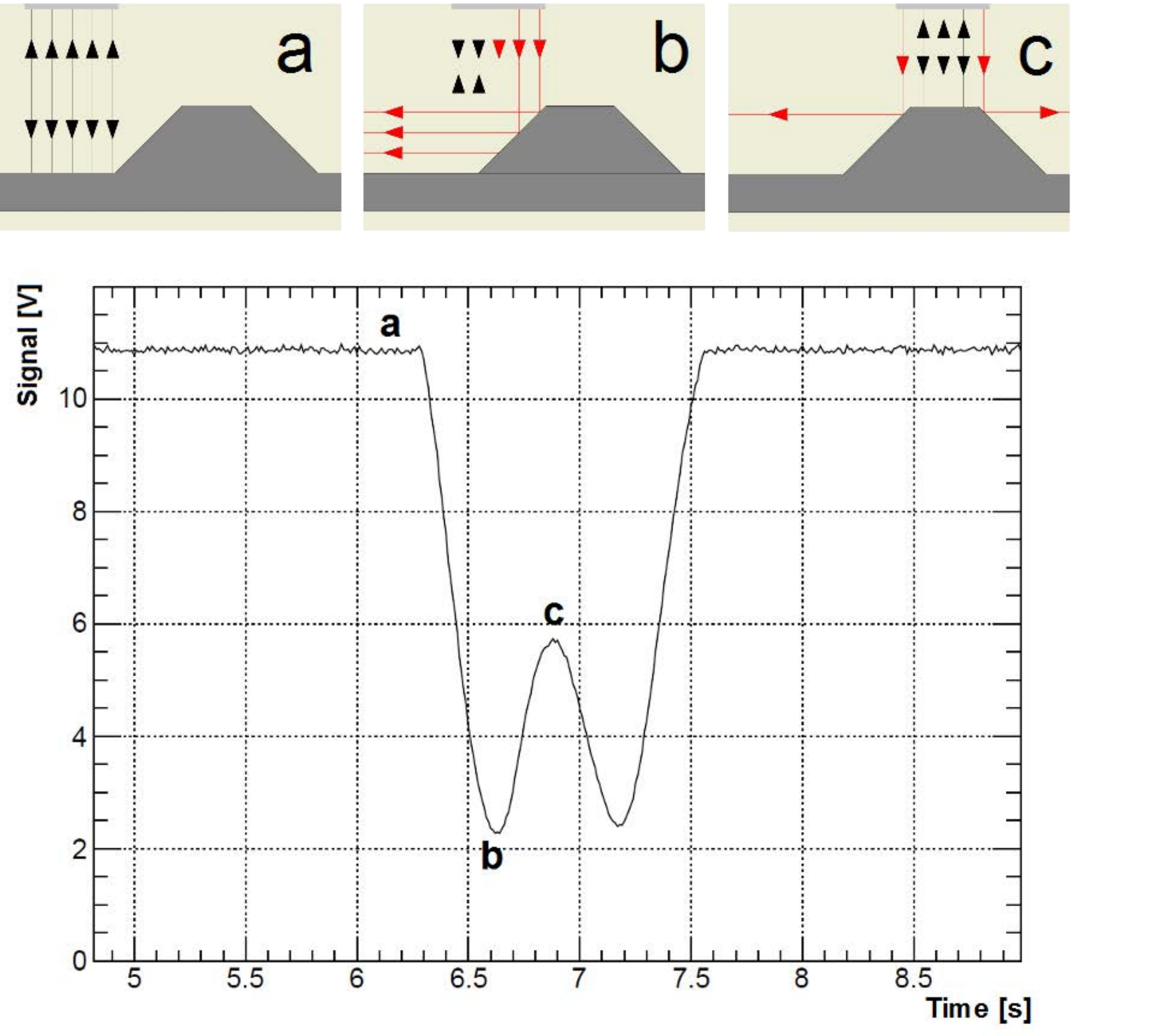}
    \caption{Functional principle of the positioning system (top) and a corresponding measured signal sequence during slide movement with constant velocity. Three representative situations are shown.}
    \label{fig:PrimaryTargetPositioningPrinciple}
\end{figure}
\section{The infrared light-based positioning system}

To monitor the slide position, light is injected perpendicular
on a reflector by a bidirectional light guide. In this way, the sensor and electronics can be placed at a safe distance --- up to several meters --- from the interaction point of the antiprotons. The light intensity collected via the light guide depends on its distance to the reflector and on the surface characteristics of the reflector as illustrated in the top part of \autoref{fig:PrimaryTargetPositioningPrinciple}. The signal sequence measured by a sensor while moving the reflector with constant velocity over the light guide is shown in the bottom part of the figure. In this study, time is a measure of the position of the reflector with respect to the light guide. In situation (a),  the incident light hits the reflector (gray) perpendicularly, and the largest intensity is reflected towards the light guide (black arrows). This results in a maximum of the measured voltage signal as shown in the bottom part of the figure. In position (b), the surface reflects a fraction of the incident light out of the acceptance of the light guide (red arrows). A suitable reflector geometry produces a minimum in the voltage signal at this position. While reaching the top of the elevation the signal rises again and forms a local maximum (c). Following this, the signal sequence is reversed.
Both minima are easy to identify and can be used for positioning the target system. The height of the local maximum however, depends strongly on the alignment of the plateau of the ridge with respect to the light guide. As a consequence, the signal shape and amplitude will vary with each modification of the light guide setup.

\section{The prototype setup}
\label{sec:prototype}

\begin{figure}[htbp]
    \centering
    \includegraphics[width=\columnwidth]{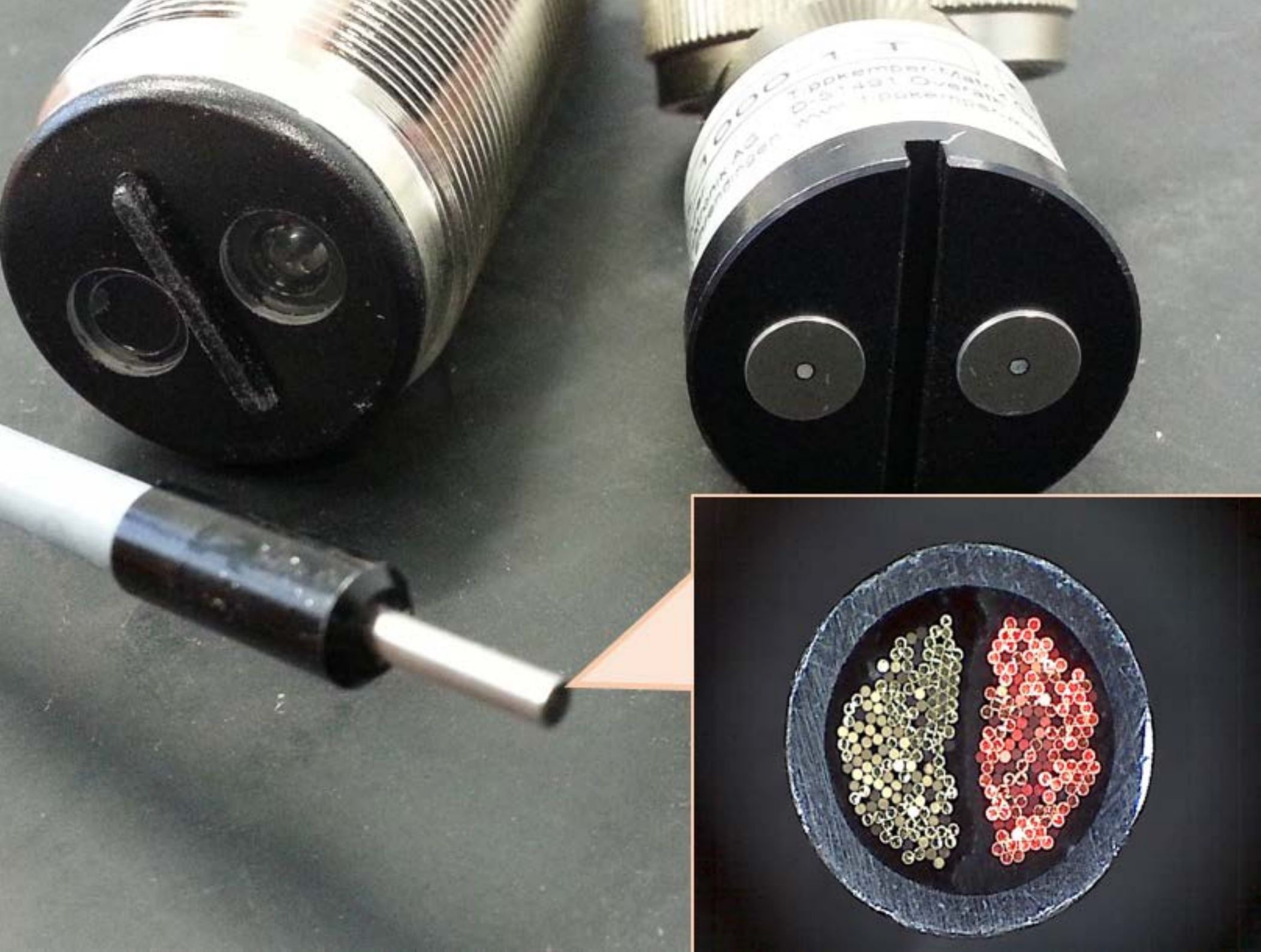}
    \caption{Photographs of the Tippkemper-Matrix IRS-U2LA optical sensor and the bisected light guide. Its two sections were illuminated differently for demonstration.}
    \label{fig:LightGuide_Zones}
\end{figure}

\begin{figure}[htbp]
	\centering
	\includegraphics[width=\columnwidth]{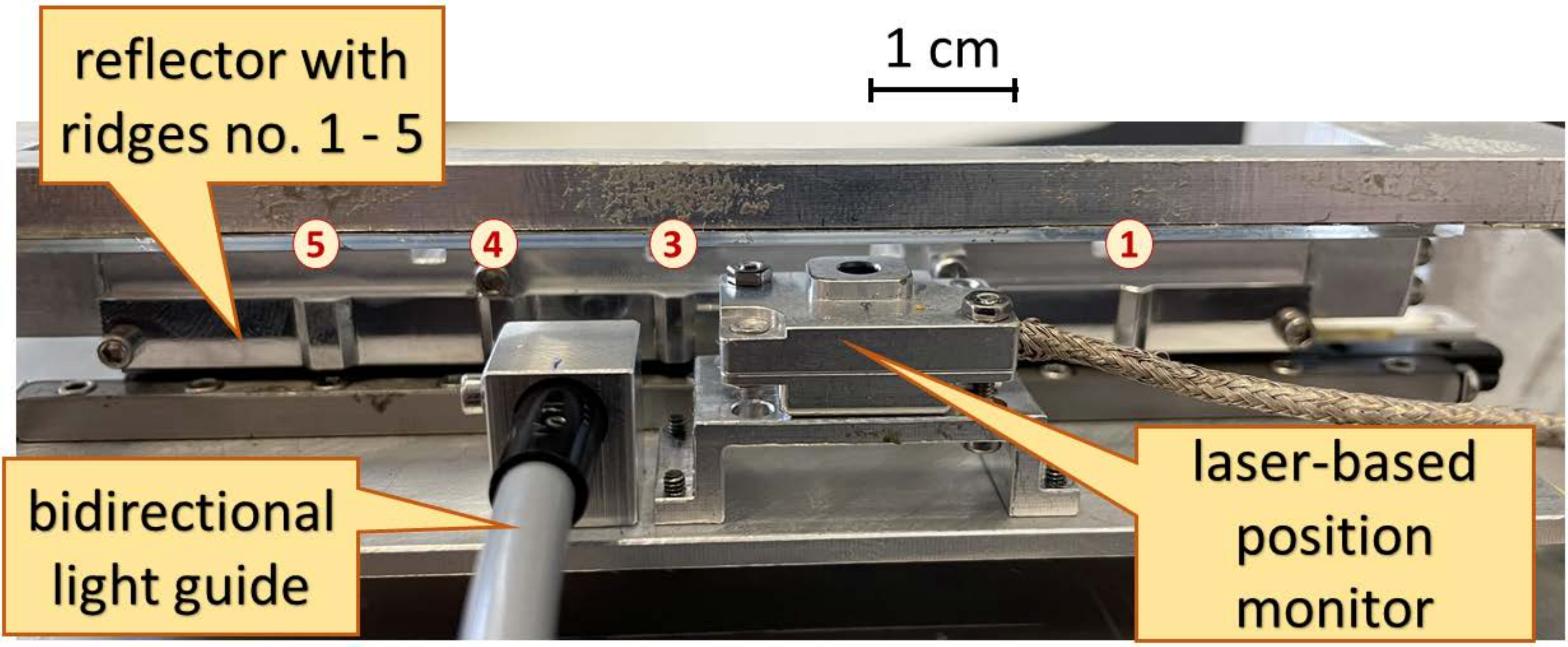}
	\caption{Rear view of the positioning stage with the light guide seen on the left and the laser-based position monitor in the center. In the background the reflector with different vertical elevations is visible.}
	\label{fig:PrimaryTargetPositioningSystem}
\end{figure}

In a prototype setup we used a Tippkemper-Matrix opto-electronic analog sensor of type IRS-U-2LA \cite{Tippkemper} using infrared light with \SI{880}{nm}, coupled to a \SI{1}{m} long bisected light guide with a diameter of \SI{2}{mm} (up to \SI{15}{m} length available) and an optical aperture angle of the light guide's head of approximately \SI{65}{\degree}.

Both ends of the light guide are shown to the left in \autoref{fig:LightGuide_Zones}. The insert shows an enlarged view of both sections of the light guide. For this photograph the two sections were illuminated differently.
The light guide is coupled to the laser and the optical sensor, respectively, by means of a proximity sensor, which is depicted in the upper right of \autoref{fig:LightGuide_Zones}.

\begin{figure}[htbp]
    \centering
    \includegraphics[width=\columnwidth]{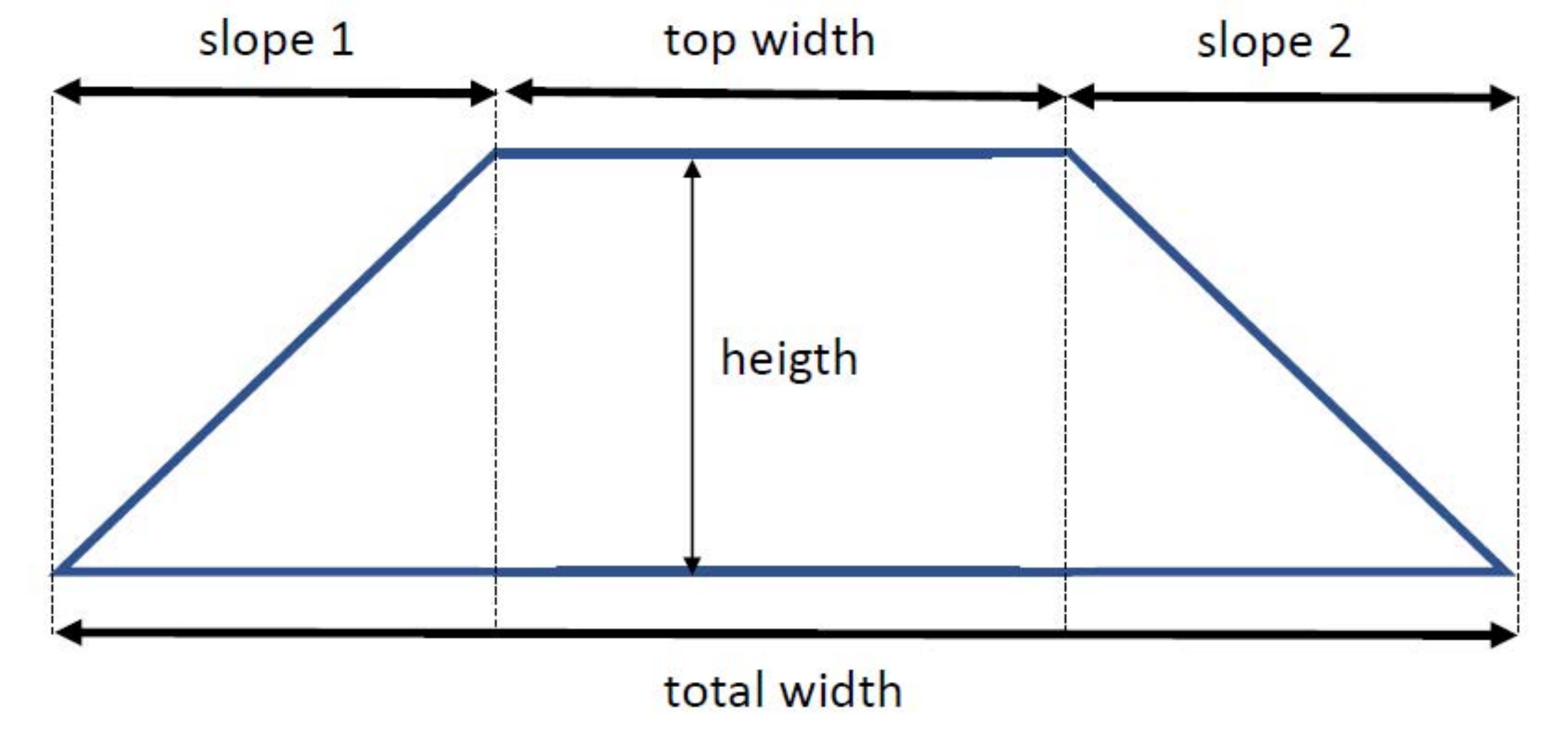}
    \vspace{0.2cm}\\
    \includegraphics[width=0.95\columnwidth]{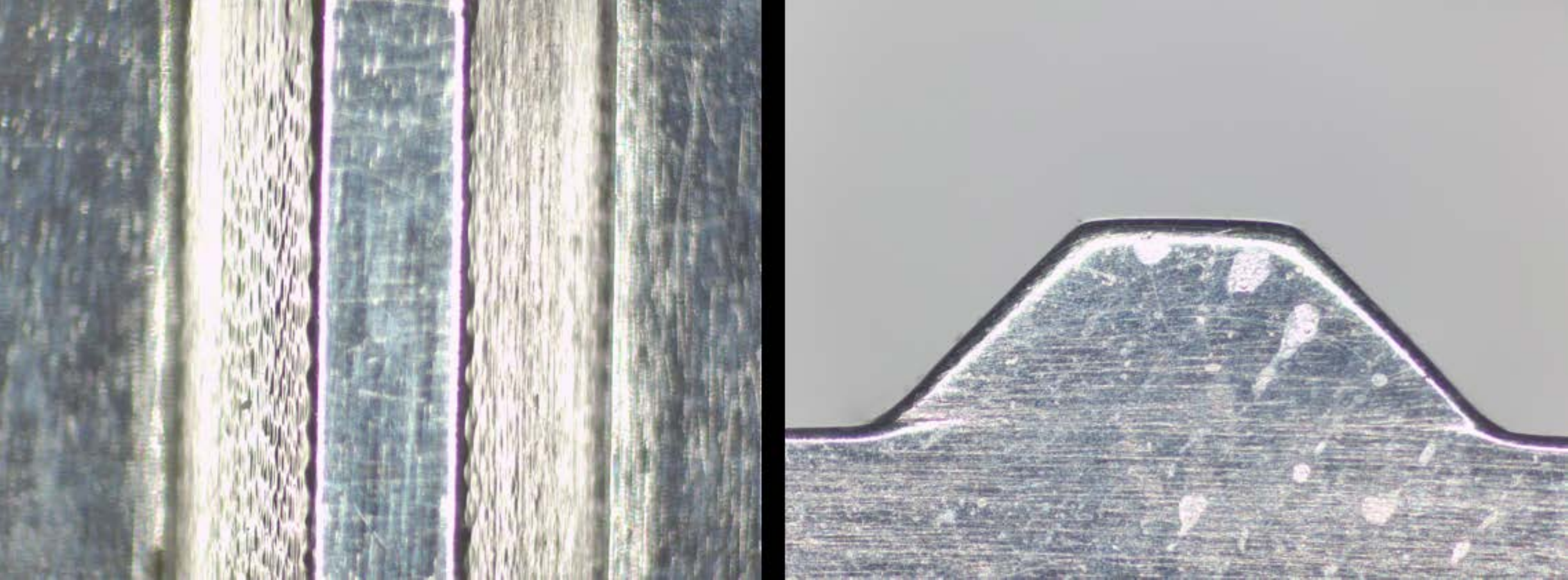}
    \caption{
    Top: parameters describing the ridge geometry (c.f.\ \autoref{tab:geometry}). Bottom: microscope photographs of ridge no.\,2 viewed from the top (left) and from the side
    (right).}
\label{fig:ridge}
\end{figure}

\begin{table}[htbp]
    \centering
    \begin{tabular}{ccccccc}
        \hline
         \multirow{2}{*}{No.} & \multicolumn{2}{c}{Total width} & \multicolumn{2}{c}{Top width} & Slope 1 & Slope 2 \\
        & Design & Actual & Design & Actual & Actual & Actual \\ \hline
        1 & 2.5 & 2.5 & 0.5 & 0.5 & 1.0 & 1.0 \\
        2 & 3.0 & 2.9 & 1.0 & 1.0 & 0.9 & 1.0 \\
        3 & 3.5 & 3.4 & 1.5 & 1.5 & 1.0 & 1.0 \\
        4 & 4.0 & 3.9 & 2.0 & 2.1 & 1.0 & 0.8 \\
        5 & 4.5 & 4.5 & 2.5 & 2.3 & 1.2 & 1.0 \\ \hline
    \end{tabular}
    \caption{Design and actual values in \SI{}{\mm} for the five different ridges of the reflector. All values were measured using a ZEISS O-INSPECT 5/4/3 \cite{ZEISSOI}. The uncertainty of the top width is about \SI{0.1}{\mm} for ridge no.\,1--4 and \SI{0.2}{\mm} for no.\,5. Each ridge had a height of approximately \SI{1}{\mm}.}
    \label{tab:geometry}
\end{table}

\begin{figure}[htbp]
    \centering
    \includegraphics[width=0.9\textwidth]{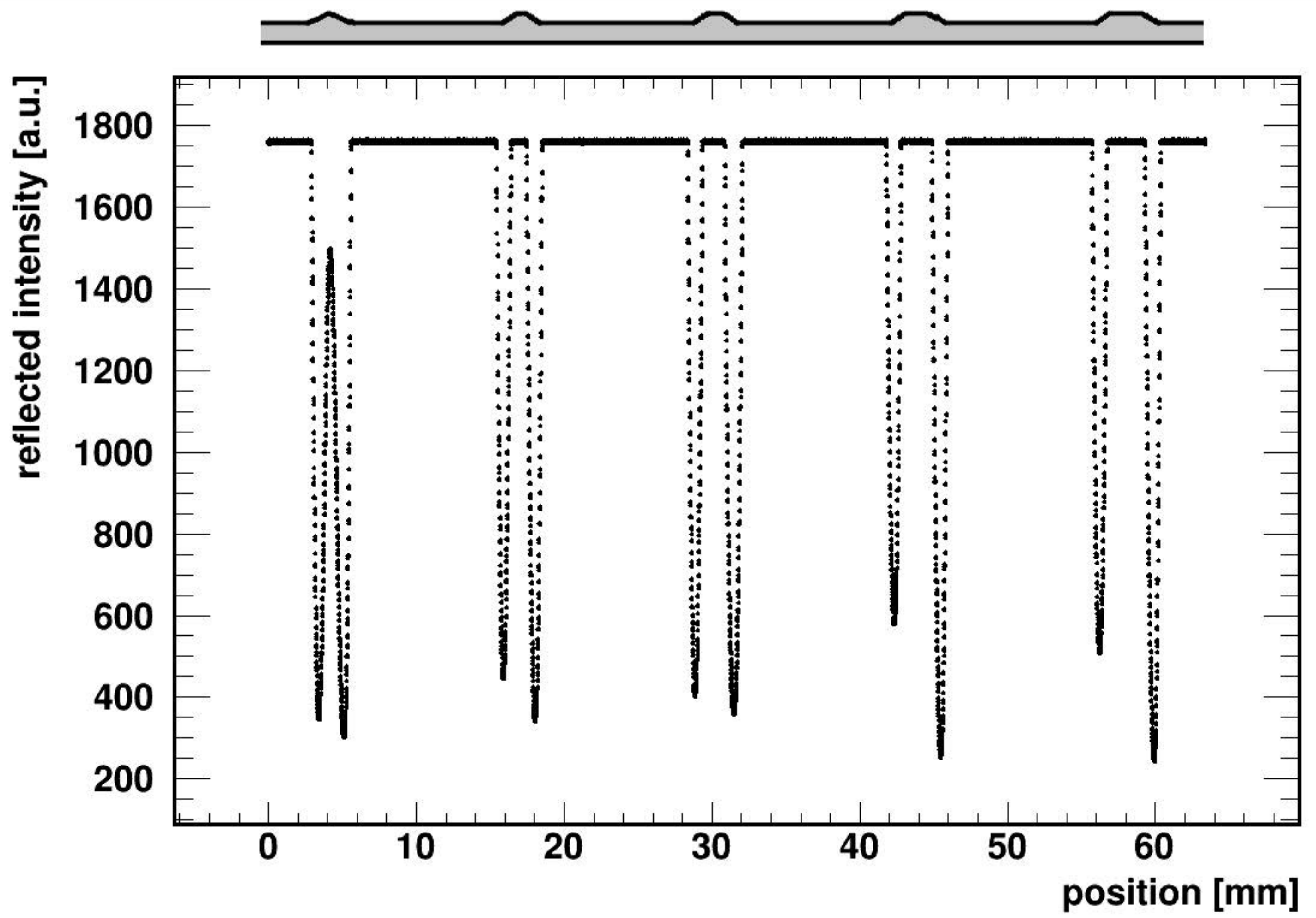}
    \caption{Measured signal of the Tippkemper sensor for one pass of the setup over the entire reflector at a distance of \SI{4}{\mm}. Except for the pair on the left, the signal saturated between the two related minima. Therefore, unlike in \autoref{fig:PrimaryTargetPositioningPrinciple}, the position of local maxima could not be observed. The distance between the minima correlates with the plateau sizes listed in \autoref{tab:geometry}.
    The top part illustrates the geometry of corresponding ridges nos.\,1 to 5 from left to right.}
    \label{fig:segment}
\end{figure}

This system was mounted on the target setup actuated by a friction-based piezo motor (Piezo LEGS Linear 6N, LL1011D \cite{PiezoMotor}).
To independently verify the position, a laser-based position monitor (MicroE Systems M1000V \cite{Encoder}) was mounted together with the light guide. The encoder of that position monitor has a native resolution of \SI{5}{\micro\metre} when used without interpolation. However, since this encoder is not radiation hard, it cannot be used during operation at {\panda}.
A photography of the prototype setup is shown in \autoref{fig:PrimaryTargetPositioningSystem}.
On the left, the bi-directional light guide can be seen. The laser-based position monitor is located in the center. In the background, the moving reflector with five vertical ridges is visible. At \panda, the  reflector will be mounted on the $z$-position slide, see \autoref{fig:setup}.

The prototype setup was controlled with Arduino Due microcontrollers using the onboard analog-to-digital converters for the measurements. One microcontroller was used exclusively to process the signal of the laser-based position monitor. The second microcontroller served as a control system, measuring the signal of the Tippkemper sensor, controlling the movements with the piezo motors, and to save the recorded data of both sensors.

A milled and surface treated aluminium plate (surface roughness $Ra=$ \SI{0.4}{\micro\m}) with five different trapezoidal shaped ridges was used as reflector. The parameters describing the geometry of the ridges are defined in the top part of \autoref{fig:ridge}.
In \autoref{tab:geometry}, the design values of the 5 ridges on the reflector are listed, as are the actual values of the produced reflector, measured with a Zeiss O-Inspect 543. The width of the top of the shapes was varied in the range between \SI{0,5}{\mm} and \SI{2,5}{\mm}.
To verify the surface quality, photographs were taken with a microscope, showing the aluminium surface with only minor irregularities and well defined edges. As an example, the lower part of \autoref{fig:ridge} shows photographs of ridge no.\,2.

Since the light guide was bisected in two distinct sectors, see \autoref{fig:LightGuide_Zones}, care had to be taken of the light guide's alignment:
defined as the angle $\phi$ between the separation line of the two sectors of the light guide and the direction of movement of the reflector.  The positive value of $\phi$ means that the emitting sector approaches the ridge first.
The head of the light guide was placed in a bracket in front of the aluminium reflector so that the incident light is perpendicular to the reflector surface, while still allowing to vary distance and alignment. Even after carefully inserting the light guide, a deviation of $\Delta\phi= \pm \ang{15}$ from its intended direction can not be excluded.

In the following, the gap between the light guide and the flat top of the ridges will be denoted as $d_\mathrm{LR}$. Due to the limited space available in the prototype, the gap is limited to a ranges of $d_\mathrm{LR}$ = 0 to \SI{5}{\mm}. In practice, a certain minimum distance of about \SI{1}{\mm} is required for the reflected light to be able to reach the receiving sector of the light guide.

\section{Systematic study of the positioning system}
%
\begin{figure}[htbp]
    \centering
    \includegraphics[width=0.70\textwidth]{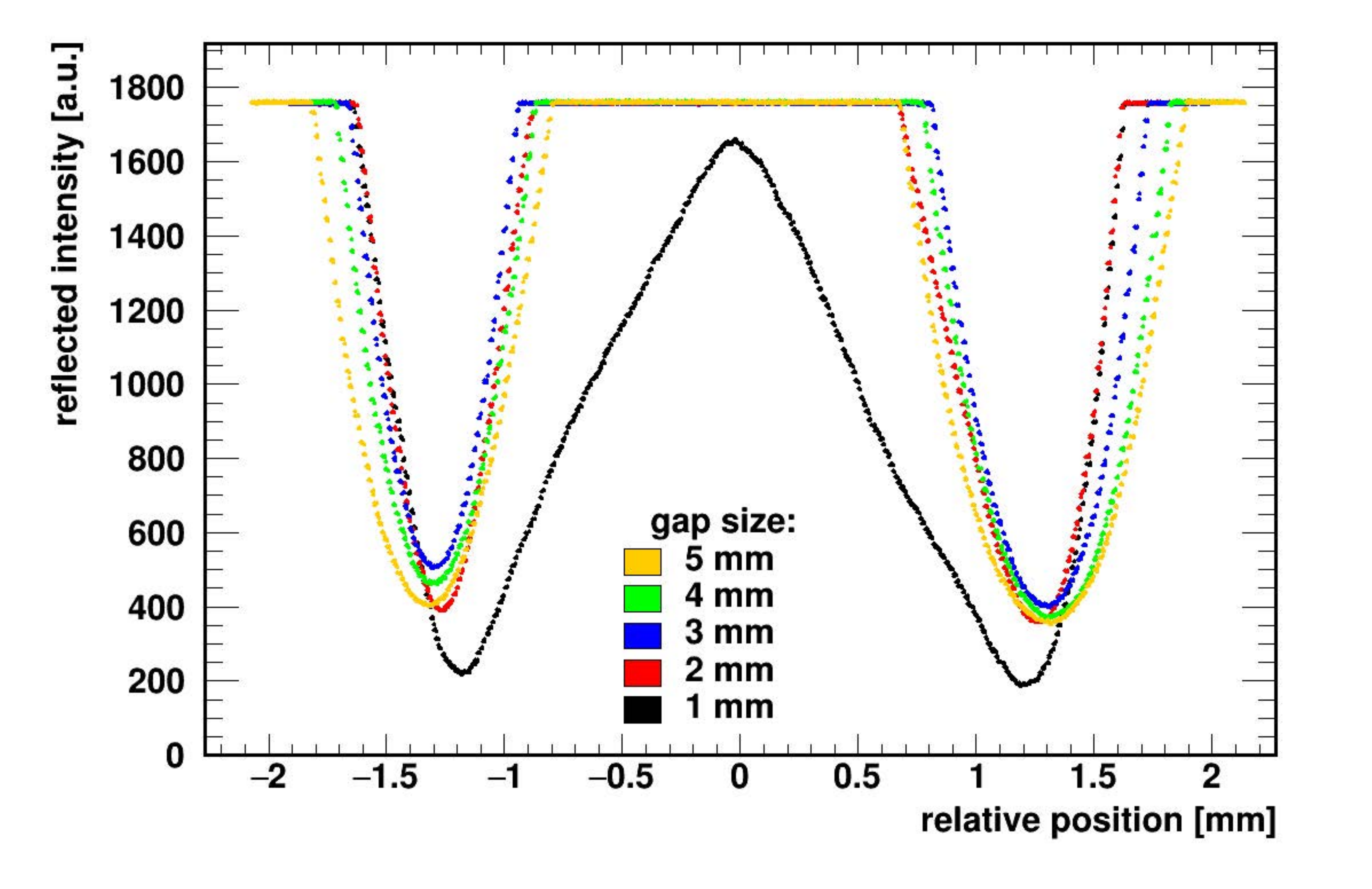}
    \caption{Signal of ridge no.\,3 measured at different gaps $d_\mathrm{LR}$. The relative positions were chosen in such a way that the midpoint between the two minima lies on top of each other.}
    \label{fig:signal}
\end{figure}
In the following, the detected light intensity is studied when changing the gap $d_\mathrm{LR}$ between the light guide and the reflector as well as the orientation $\phi$ of the bisected light guide. The default value for the gap was $d_\mathrm{LR}=$ \SI{2}{\mm} and the parting line between the two sections was perpendicular to the ridges, i.e.\ $\phi= \ang{0}$.

The lower part of \autoref{fig:segment} shows the measured signal of the Tippkemper sensor for one full pass of the setup over the entire reflector at a gap of $d_\mathrm{LR}=$ \SI{4}{\mm}. If the injected light hits the slopes of the ridges it is scattered out of the acceptance of the light guide. The five ridges of the reflector have very similar slopes and differ mainly in the width of the plateau, see \autoref{tab:geometry}. As a consequence, the depths of the minima in the detected light are similar (c.f.\ situation (b) in \autoref{fig:PrimaryTargetPositioningPrinciple}).
Because of a high incident light intensity, the sensor reached saturation of the analog-to-digital converter at positions away from the ridges and in the center of the ridges where a significant fraction of the injected light is reflected into the acceptance of the collecting section (c.f.\ \autoref{fig:LightGuide_Zones}).

\begin{figure}[htbp]
    \centering
    \includegraphics[width=0.90\textwidth]{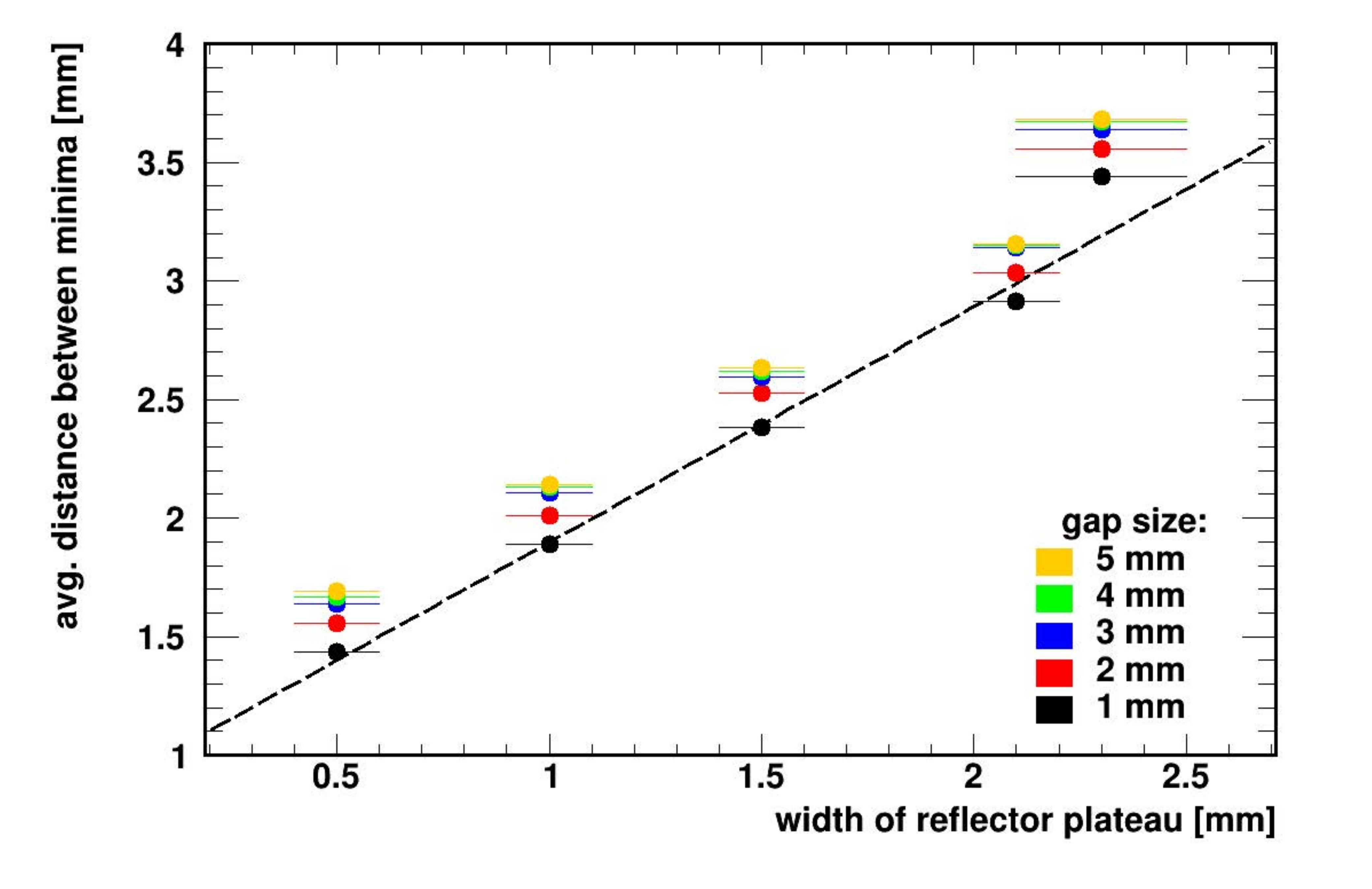}
    \caption{Mean separation between two minima for the different widths of the reflector plateaus and different gaps $d_\mathrm{LR}$. For each gap 200 scans in total (100 for each direction of motion) were performed. As an example, the regression line for $d_\mathrm{LR}=$ \SI{1}{\mm} illustrates the approximate linear dependency.}
    \label{fig:mindist}
\end{figure}

\autoref{fig:signal} shows the signal from ridge no.\,3 for different gaps between the light guide and the plateau. For all distances, the minima are clearly visible. The adjustment of the gap required an intervention into the mechanical setup. To overlay the different measurements for the different gaps, the relative positions were aligned at the midpoint between the two minima. The relative position of the minima shifted only slightly for larger gaps. Because of the bisected geometry of the light guide, a major fraction of the reflected light does not enter the sensitive area of the light guide for very small gaps $\sim$ \SI{1}{\mm}. As a consequence, the sensor did not saturate and a local maximum developed in the center between the two minima (black points in \autoref{fig:signal}).

The widths of the plateaus seen in \autoref{fig:segment} correspond to the widths of the ridge as displayed in the upper part of the figure and listed in \autoref{tab:geometry}. However, the observed plateaus also depend on the incident light intensity and hence also on the gap. In fact, the locations of the minima in \autoref{fig:segment} are less sensitive to the intensity of the incident light. We evaluated the mean distance ${\Delta}d_{min}$ between two minima from 200 scans to characterize the observed intensity function and to study the effect of the width of the ridge.

\autoref{fig:mindist} shows ${\Delta}d_{min}$ as a function of the measured top widths (c.f.\ \autoref{tab:geometry}).
The different colors correspond to different gaps. The regression line for a gap of \SI{1}{\mm} illustrates a approximately linear dependency except for the widest ridge. This irregularity may reflect the uncertainty in the optical measurements.

\section{Precision of the position determination}
%
High accuracy and high precision of the $z$-position slide is required. Since the movement can not be independently controlled, a self-calibration procedure is needed which provides a low bias and a high repeatability. To determine the quality of the prototype system, the setup was moved repeatedly over all five ridges. The analog output of the Tippkemper sensor and the position information of the laser-based monitor was recorded simultaneously. The data could than be used in an offline analysis to find all minima in the analog voltage signal and their corresponding positions. The midpoint between two corresponding minima was used as a measure of the $z$-position.

\begin{figure}[htbp]
    \centering
    \includegraphics[width=\columnwidth]{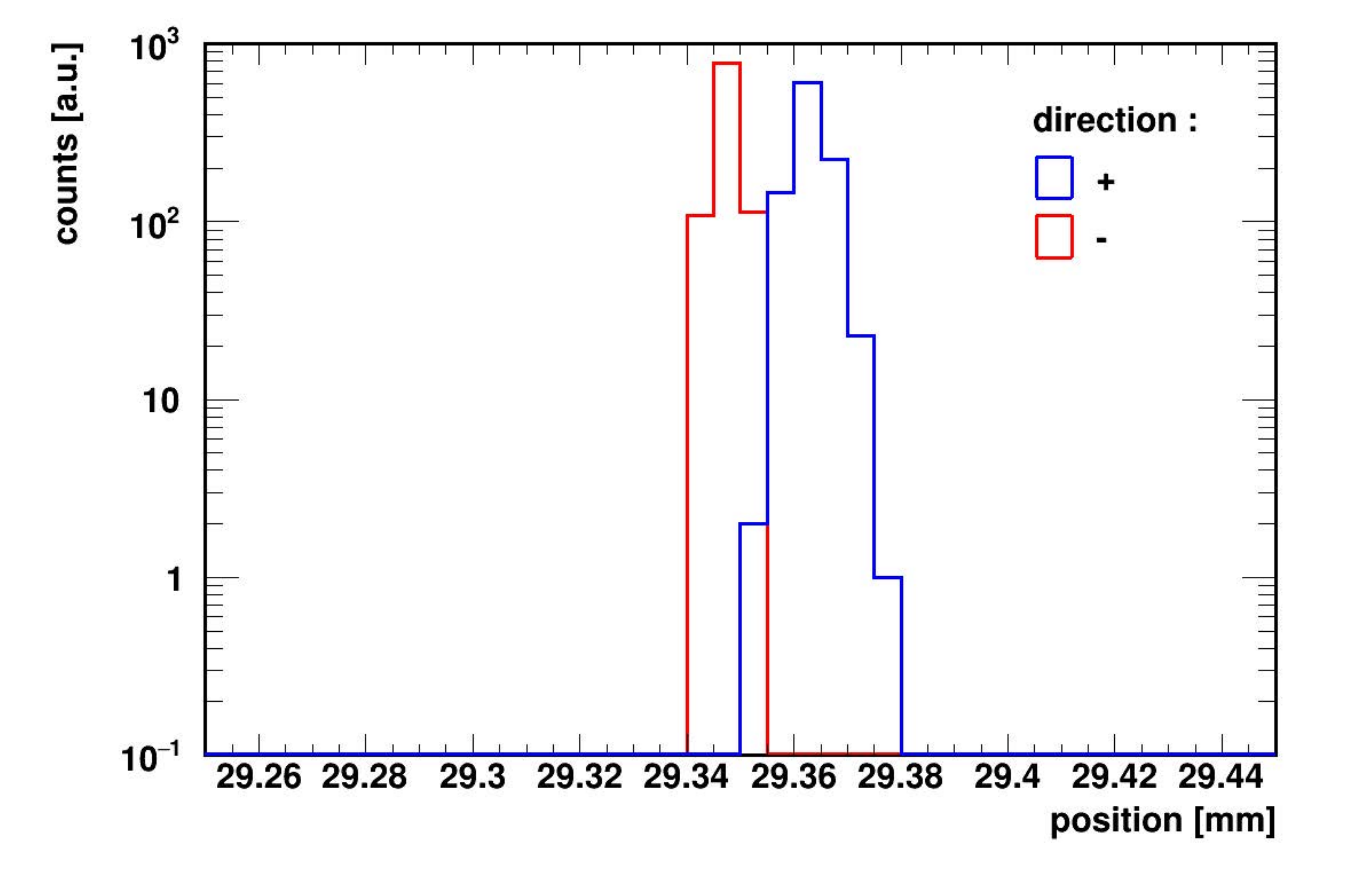}
    \caption{Distribution of the position of the first minimum for ridge no.\,3 separated for the two directions of motion. Here, a gap of $d_\mathrm{LR}=$ \SI{2}{\mm} and an orientation of the light guide of $\phi= \ang{0}$ was chosen. For each direction of the motor, 1000 scans were performed. One bin of the histogram corresponds to one step of the laser-based position monitor.}
    \label{fig:peak}
\end{figure}

As an example, \autoref{fig:peak} shows the measured position of the first peak for ridge no.\,3. For this measurement, 1000 scans at a fixed gap of \SI{2}{\mm} were performed. The data are shown separately for the two directions of motion. The measured position depends on the direction because of the finite processing time of the Tippkemper module. The time to process the optical information within the readout electronics of the sensor is specified by the manufacturer with up to \SI{5}{\ms}.
Based on the average driving speed of the setup, the difference between the two positions in \autoref{fig:peak} corresponds to approximately \SI{6.5}{\milli\s}. This difference is within the specified value when considering that the delay affects the motion in an opposite way for the two directions. Using an improved sensor with less delay could reduce this effect. In the present setup we can eliminate this bias by approaching a position always from the same side, and calibrating each position taking the delay into account. It was verified that the delay is constant over time by determining it for each pair of left$/$right movements. The root-mean-
square (rms) width of the peaks is $\sigma \sim$ \SI{5}{\micro\m}, which proves the excellent precision of the setup.

\begin{table}[htbp]
    \centering
    \begin{tabular}{ccc}
    \hline
    Ridge no. & $\sigma_\mathrm{min1}$ [\SI{}{\micro\m}] & $\sigma_\mathrm{min2}$ [\SI{}{\micro\m}]\\
    \hline
    1 & \SI{2.6 \pm 0.2}{}	& \SI{3.6 \pm 0.3}{} \\
    2 & \SI{2.5 \pm 0.2}{}	& \SI{2.6 \pm 0.2}{} \\
    3 & \SI{3.2 \pm 0.3}{}	& \SI{4.9 \pm 0.4}{} \\
    4 & \SI{2.6 \pm 0.2}{}	& \SI{2.4 \pm 0.2}{} \\
    5 & \SI{2.9 \pm 0.2}{}	& \SI{2.5 \pm 0.2}{} \\
    \hline
    \end{tabular}
    \caption{Position resolution for all 5 ridges at a fixed gap $d_\mathrm{LR}=$ \SI{2.0 \pm 0.2}{\mm} and an alignment angle of $\phi= \ang{0} \pm \ang{15}$. For this measurement, 100 scans were performed for each ridge.}
    \label{tab:prec}
\end{table}

100 scans over of all five ridges were performed. For a selected gap of $d_\mathrm{LR}=$ \SI{2}{\mm} the achieved precision is given in \autoref{tab:prec}. Since each ridge results in two minima, both rms widths are listed. For ridge no.\,3 the resolution seems to be slightly worse. In all cases, the resolution is better than \SI{5}{\micro\m}. The light guide was first installed with a parallel alignment of $\phi= \ang{0}$. Then, the light guide was rotated by \ang{90} and the measurement was repeated. Finally, the transmitting and receiving side were inverted to $\phi= \ang{-90}$.
For each of the three alignment angles $\phi$ at a fixed gap of {\SI{2}{\mm}}, the rms widths for all five ridges were determined. The weighted mean standard deviations, serving as a measure of precision for each alignment, are listed in \autoref{tab:configs}. A parallel alignment is superior to a perpendicular one. In the parallel case, the light guide is symmetric relative to the axis of motion of the reflector and combined with the symmetric reflector shape this is expected to lead to symmetric signals at the left and right slope. In the perpendicular case, the distribution of light depends on whether the slope is moved near the emitting or receiving sector first.

\begin{table}[htbp]
    \centering
    \begin{tabular}{rc}
    \hline
    \centering Alignment $\phi$ 		& Precision $\sigma$ [\SI{}{\micro\metre}] \\
    \hline
    $\ang{0} \pm \ang{15}$   & \SI{2.8 \pm 0.1}{} 	\\
    $\ang{+90} \pm \ang{15}$  & \SI{8.8 \pm 0.2}{} 	\\
    $\ang{-90} \pm \ang{15}$ & \SI{10.9 \pm 0.3}{}	\\
    \hline
    \end{tabular}
    \caption{Dependence of the measured precision for the position on the alignment of the bisected light guide. The gap was \SI{2}{\mm}. A positive angle $\phi$ indicates that the emitting sector of the light guide approached the trapezoidal shaped ridge first.}
    \label{tab:configs}
\end{table}

\begin{table}[htbp]
    \centering
    \begin{tabular}{cc}
    \hline
    Gap $d_{LR}$ [\SI{}{\milli\m}] & Precision $\sigma$ [\SI{}{\micro\m}]\\
    \hline
    \SI{1.0 \pm 0.2}{} & \SI{3.7 \pm 0.1}{} \\
    \SI{2.0 \pm 0.2}{} & \SI{2.8 \pm 0.1}{} \\
    \SI{3.0 \pm 0.2}{} & \SI{3.4 \pm 0.1}{} \\
    \SI{4.0 \pm 0.2}{} & \SI{3.8 \pm 0.1}{} \\
    \SI{5.0 \pm 0.2}{} & \SI{5.2 \pm 0.1}{} \\
    \hline
    \end{tabular}
    \caption{Dependence of the measured precision of the position on the gap between the light guide and the reflector. The alignment angle was $\ang{0} \pm \ang{15}$.}
    \label{tab:fom}
\end{table}

The dependence on the gap was also studied and the weighted mean standard deviation of all measured minima was used as a measure of precision. The results are listed in \autoref{tab:fom}. For the setup used in our study, a distance of \SI{2}{\mm} was optimal, with a significant degradation of the precision with increasing or decreasing gaps. In all cases, the precision was better than $\sigma \approx$ \SI{5}{\micro\m}.

In our test setup, the accuracy of the absolute position determination was limited by the resolution of \SI{5}{\micro\metre} of the laser-based encoder. Any further improvement of the accuracy would require the use of the encoder's interpolation feature.

To quantify the reliability with which a selected position can be approached, we determined the maximum deviation from the mean position for 1000 scans for ridge no.\,3 as shown in \autoref{fig:peak}. The base widths are \SI{11}{\micro\m} and \SI{25}{\micro\m} for the two directions. Approaching a position from the same direction, the largest deviation will be of the order of $\pm$ \SI{15}{\micro\m}.

During a long-term measurement, the five ridges were scanned 7000 times for each direction of motion. For the two directions, maximum deviations of {\SI{35.4}{\micro\m}} and {\SI{30.4}{\micro\m}} were observed. Thus, the repeatability was better than $\pm$\SI{18}{\micro\m}.

\section{Summary}
%
A positioning sytem was built using a bisected light guide with infrared light and a low-priced readout system based on microcontrollers. The system was proven to operate with a resolution of better than \SI{5}{\micro\m}, which is more than sufficient for use in the hypernuclear and hyperatom experiments at \panda. In
14\,000 measurements, the repeatability was better than $\pm$ \SI{18}{\micro\m} and thus far better than the maximum allowed deviation at \panda of $\simeq$ \SI{300}{\micro\m}.

Finally, it is important to note, that in case of a loss of the sensor status, e.g.\ during long-time operation at \panda, the position can be re-calibrated by approaching the endpoints or by a scan over one or more ridges with an accuracy of better than
$\pm$ \SI{5}{\micro\m}. We, therefore, conclude that the reliability of the system will be sufficient for a safe operation at \panda.

In future, several questions could be addressed
for a further improvement of such a system.
\begin{itemize}
    \item
    Alternative geometries of the ridges, e.g.\ different slopes or different shapes, could be investigated.
    \item
    Polishing or gold-plating the reflector could reduce local variations of the reflectivity.
    \item
    Improved sensor electronics could help to minimize the latency of the readout.
    \item
    Using a statistically mixed light guide could reduce the sensitivity to the direction of movement.
    \item
    The placement of the light guide relative to the reflector surface could be optimized.
\end{itemize}

Commercially available solutions such as laser-based position encoders may offer better resolutions, however, the major advantage of the presented system is its endurance under harsh environmental conditions. The presented method could also be of interest for other applications.

\section*{Acknowledgement}
This project has received travel funds from the European Union’s Horizon 2020 research and innovation programme under grant agreement No 824093.

We would also like to thank Heinrich Leithoff from the Helmholtz Institute Mainz for his contribution to this work by measuring the reflector geometry with a Zeiss O-Inspect.

The presented data were in part collected within the work of the diploma thesis of Falk Schupp and the doctoral thesis of Michael B\"olting, both at the Helmholtz Institute Mainz of the Johannes Gutenberg University Mainz.

\bibliography{Schupp_Target.bib}

\end{document}